\documentclass[superscriptaddress, reprint,nofootinbib]{revtex4-2}
\usepackage{graphicx}
\graphicspath{ {../fig/}}

\usepackage{amssymb}
\usepackage{amsmath}
\usepackage{amsthm}
\usepackage{amsfonts}
\usepackage{xfrac}
\usepackage{wasysym}
\usepackage{textcomp}
\usepackage{bbm}

\usepackage{braket}
\usepackage[version = 4]{mhchem}
\usepackage{epstopdf} 
\usepackage{physics}

\usepackage{xcolor}

\usepackage{hyperref}
\usepackage{array}
\usepackage{comment}

\newcommand{\figSize}{0.95}

\begin{document}

\author{N. P. Giha}
\email{giha@umich.edu}

\affiliation{Department of Nuclear Engineering and Radiological Sciences, University of Michigan, Ann Arbor, MI 48109, USA}
\affiliation{Physics Division, Argonne National Laboratory, Lemont, IL 60439, USA}

\author{S. Marin}
\affiliation{Department of Nuclear Engineering and Radiological Sciences, University of Michigan, Ann Arbor, MI 48109, USA}

\author{I. A. Tolstukhin}
\email{itolstukhin@anl.gov}

\author{M. B. Oberling}
\author{R. A. Knaack}
\author{C. Mueller-Gatermann}

\affiliation{Physics Division, Argonne National Laboratory, Lemont, IL 60439, USA}

\author{A. Korichi}
\affiliation{IJCLab, IN2P3-CNRS, Université Paris-Saclay, bat 104-108, Orsay Campus, 91405, Orsay, France}
\affiliation{Physics Division, Argonne National Laboratory, Lemont, IL 60439, USA}

\author{K. Bhatt}
\author{M. P. Carpenter}
\author{C. Fougères}
\author{V. Karayonchev}
\author{B. P. Kay}
\author{T. Lauritsen}
\author{D. Seweryniak}
\author{N. Watwood}
\affiliation{Physics Division, Argonne National Laboratory, Lemont, IL 60439, USA}

\author{D. L. Duke}
\author{S. Mosby}
\author{K. B. Montoya}
\author{D. S. Connolly}
\affiliation{Los Alamos National Laboratory, Los Alamos, New Mexico 87545, US}

\author{W. Loveland}
\affiliation{Chemistry Department, Oregon State University, Corvallis, OR 97331, USA}

\author{I. E. Hernandez}

\author{S. D. Clarke}
\affiliation{Department of Nuclear Engineering and Radiological Sciences, University of Michigan, Ann Arbor, MI 48109, USA}

\author{S. A. Pozzi}
\affiliation{Department of Nuclear Engineering and Radiological Sciences, University of Michigan, Ann Arbor, MI 48109, USA}
\affiliation{Department of Physics, University of Michigan, Ann Arbor, MI 48109, USA}

\author{F. Tovesson}
\affiliation{Physics Division, Argonne National Laboratory, Lemont, IL 60439, USA}

\title{Measurement of spin vs. TKE of \textsuperscript{144}Ba produced in spontaneous fission of \textsuperscript{252}Cf}

\date{\today}

\begin{abstract}
We measure the average spin of $^{144}$Ba, a common fragment produced in $^{252}$Cf(sf), as a function of the total kinetic energy (TKE). We combined for the first time a twin Frisch-gridded ionization chamber with a world-class $\gamma$-ray spectrometer that was designed to measure high-multiplicity $\gamma$-ray events, Gammasphere. The chamber, loaded with a $^{252}$Cf(sf) source, provides a fission trigger, the TKE of the fragments, the approximate fragment masses, and the polar angle of the fission axis. Gammasphere provides the total $\gamma$-ray yield, fragment identification through the tagging of decay $\gamma$ rays, and the feeding of rotational bands in the fragments. We determine the dependence of the average spin of $^{144}$Ba on the fragments’ TKE by correlating the fragment properties with the distribution of discrete levels that are fed. We find that the average spin only changes by about $0.5$~$\hbar$ across the TKE range of 158-203 MeV. The virtual independence of the spin on TKE suggests that spin is not solely generated through the statistical excitation of rotational modes, and more complex mechanisms are required.

\end{abstract}

\keywords{angular momentum of fission fragments; fission fragment de-excitation, experimental fission physics}

\maketitle


\section{Introduction}
\label{sec:Introduction}
Despite over eighty years of study since nuclear fission was discovered~\cite{Hahn1939, MEITNER1939}, important details of this nuclear complex process remain a mystery. This lack of knowledge presents an obstacle to constructing a complete predictive model for the fission process, which would aid in the modeling of fission recycling in \textit{r}-process of nucleosynthesis~\cite{Goriely2015, Vassh2019, Mumpower2020,  Vassh2020, Wang2020}, Generation-IV fast-fission reactors~\cite{Rimpault2012}, and simulating nuclear nonproliferation and safeguards scenarios~\cite{Kolos2022}. Chief among the open questions is the mechanism by which the fissioning system, often possessing little or no spin, produces fragments with spins of potentially tens of $\hbar$. Understanding the mechanism of spin generation is not only important to explain and predict $\gamma$-ray production data, but it would also shed light on the microscopic and quantum mechanical aspects of the fission dynamics.

Advances in computational models~\cite{Talou2021, Litaize2015, Vogt2009, Bulgac_FFIntrinSpins2021, Marevic2021} and experimental techniques~\cite{Leoni2022} have brought the question of spin generation to the forefront; recently a flurry of theoretical~\cite{Bertsch2019, Vogt2021, Bulgac_FFIntrinSpins2021, RandrupVogt_GenOfFragAngMom2021, Marevic2021, Stetcu2021, Randrup2022, Dossing2024} and experimental~\cite{Travar2021, Wilson2021, Gjestvang2021, Giha2023, Marin2024} work has discussed or attempted to constrain possible mechanisms. Several mechanisms have been proposed, including statistical excitation of rotational modes based on fragment temperature~\cite{Randrup2021, Talou2021}, deformation of the fragments during the fission process~\cite{Marevic2021, Bulgac_FFIntrinSpins2021}, and the Coulomb torque between misaligned fragments~\cite{bertsch2019reorientation,Scamps2022, Randrup2023_coulomb}. The models built on these descriptions often do well in reproducing integral quantities related to spin, such as average $\gamma$-ray multiplicity. To help differentiate between them, the community has focused on measuring correlations between fission observables such as fragment mass, kinetic or intrinsic energy, and $n/\gamma$-ray emission~\cite{Wilhelmy1972, Gook2014, Gjestvang2021, Travar2021, Wilson2021}.

However, no previous experiment has measured the spin distribution of a fragment as a function of total kinetic energy (TKE) of the fragments. In this work, we investigate correlations between the spin and energy generated in fission for $^{144}$Ba, a fragment commonly produced in $^{252}$Cf(sf) and fission of other actinides. We simultaneously measure prompt fission $\gamma$ rays\textemdash believed to carry most of a fission fragment's spin\textemdash and TKE. Leveraging the high-resolution, high-granularity spectroscopic capabilities of Gammasphere, we measure the intensities of known transitions and infer the relative initial population of several low-lying discrete nuclear levels in $^{144}$Ba, as a function of TKE. 

\begin{figure}[htb!]
    \centering
    \includegraphics[width = \figSize\columnwidth]{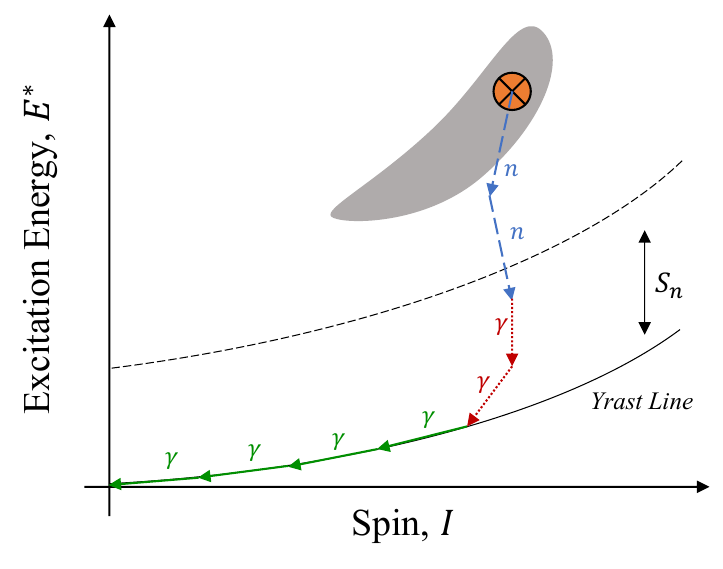}
    \caption{De-excitation of a fission fragment by emission of neutrons, statistical $\gamma$ rays, and discrete $\gamma$ rays. The yrast line follows the nuclear levels with minimum $E^*$ for a given $I$. Neutrons are in dashed blue, statistical $\gamma$ rays in dotted red, and discrete $\gamma$ rays in solid green arrows.}
    \label{fig:deexcitation}
\end{figure}

In a simplified picture of fission fragment de-excitation (Fig.~\ref{fig:deexcitation}), a nascent fragment decays in three steps. While the excitation energy of a fragment, $E^*$, exceeds the neutron separation energy ($\approx 6-8$ MeV), the fragment emits neutrons. Due to their large binding energy, neutrons remove the bulk of the fragment $E^*$ and it is commonly assumed that they remove, on average, little spin. This assumption has been called into question by a recent theoretical study~\cite{Stetcu2021}, but experimental evidence indicates neutron emission is isotropic in the reference frame of the fragment, suggesting $s$-wave emission~\cite{Gook2014}. After neutron emission, the nucleus is likely left with several MeV of $E^*$ where the nuclear level density is still high, and decays between levels in the quasi-continuum proceed by statistical $\gamma$-ray emission. Recent experimental evidence~\cite{Marin2022} suggests that the angular distribution of these statistical $E1$ $\gamma$ rays from the quasi-continuum is uncorrelated with the fission axis, so it is assumed here that they do not remove much fragment spin on average.

The fragment eventually reaches a regime of low level-density and decays along the ``yrast line'' by mainly stretched electric quadrupole ($E2$) transitions. These yrast-band transitions connect discrete nuclear levels that are well-characterized in the evalulated nuclear structure data file (ENSDF) database~\cite{ensdfSite} with deduced excitation energies and spins. This experiment was designed to leverage the well-known low-lying nuclear structure of the fragments, through $\gamma$-ray spectroscopy, in order to reconstruct the spin distribution of the fragment after statistical emission~\cite{Wilhelmy1972, Wilson2021}.

In Section~\ref{sec:Experiment}, we describe the experimental setup and procedure. Section~\ref{sec:Analysis} outlines the method for extracting the fragment spin distribution as a function of TKE. In Section~\ref{sec:Results}, we present the average spin of $^{144}$Ba as a function of TKE. We interpret the results in Section~\ref{sec:Discussion}, then conclude in Section~\ref{sec:Conclusion} and propose future analyses using these data.

\section{Experimental setup and procedure}
\label{sec:Experiment}
\begin{figure}[htb!]
    \centering
    \includegraphics[width = \figSize\columnwidth]{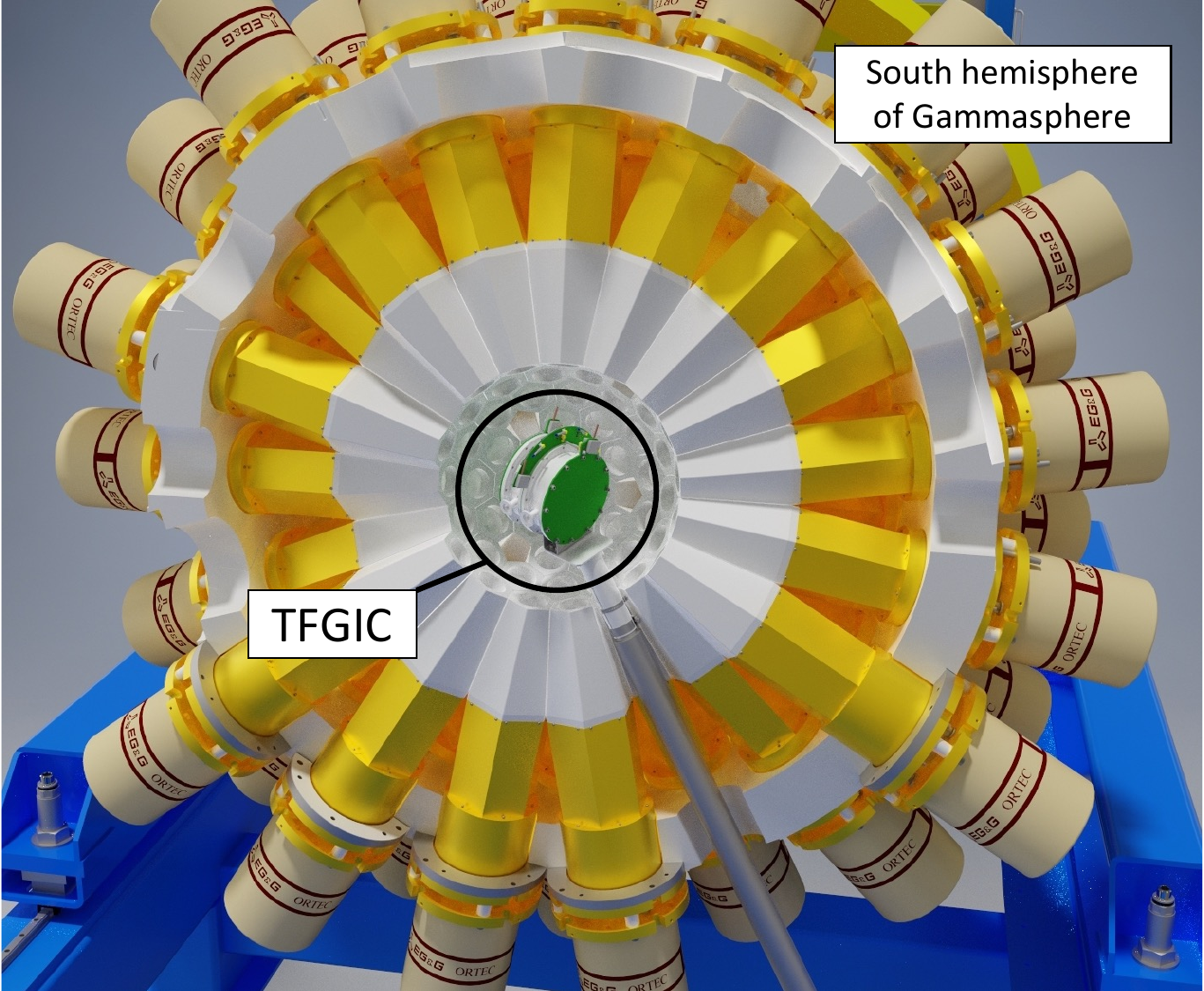}
    \caption{Rendering of the twin Frisch-gridded ionization chamber detector (TFGIC) mounted in the center of Gammasphere~\cite{Lee1990}. Only the south hemisphere is shown; the north hemisphere closes so that the TFGIC is completely surrounded.}
    \label{fig:setup}
\end{figure}

For the experiment, we combined a twin Frisch-gridded ionization chamber with Gammasphere to measure both fragments and $\gamma$ rays. The twin Frisch-gridded ionization chamber (TFGIC) at Argonne National Laboratory (ANL)~\cite{Marin2023} consists of two identical volumes enclosed between the central cathode plate and the two anode plates. The inner diameter of the chamber is 140 mm, and the distance between the cathode and anode boards is 47 mm. Frisch grids are positioned between the cathode and each anode at a distance of 7 mm from the anode. The TFGIC was loaded with a $\sim 4,000 \text{ fissions/s}$ $^{252}$Cf spontaneous fission source deposited on a 100 $\mu$g/cm$^2$ carbon foil and placed inside the Compton-suppressed high-purity germanium (HPGe) detector array Gammasphere~\cite{Lee1990} at the Argonne Tandem Linear Accelerator System (ATLAS) Facility at ANL. A rendering of the experimental setup is shown in Fig.~\ref{fig:setup}. The setup enabled simultaneous measurement of (i) both pre-neutron fission fragment masses $A_{L,H}$, kinetic energies $KE_{L,H}$, and the fission axis orientation with respect to TFGIC axis $\cos\theta_\text{f}$ and (ii) the energies $E_\gamma$ and angles of prompt $\gamma$ rays emitted following fission. The TFGIC was centered inside Gammasphere with a laser alignment gauge such that the target was at the center of Gammasphere.

The signals from the Gammasphere detectors were read out with its data acquisition system (DAQ)~\cite{gsdaq} in triggerless mode, wherein all events which surpass the discriminator of a given channel are written to disk in list mode. The GEBSort code~\cite{gebsort}, developed at Argonne, was used to merge the data from the the Gammasphere DAQ, calibrate and perform pole-zero corrections, and form coincidence events. We modified the source code to output a \textsc{root}~\cite{root} tree of these events.

All TFGIC electrode waveforms were digitized with a CAEN V1740D digitizer~\cite{caenV1740D} and stored on disk. The two anode signals from each side of the TFGIC were cloned with a CAEN N625 fan-in fan-out unit~\cite{caenN625} and routed to two empty digitizer channels in the Gammasphere DAQ to provide a coincidence fission trigger in post-processing. A suite of codes mentioned in Ref.~\cite{Marin2023} was used to analyze the waveforms and apply the $2E$ method~\cite{Duke2016}, producing a \textsc{root}~\cite{root} tree of fission fragment events.

To synchronize the separate digital DAQ systems, we propagated a 50-MHz clock from the Gammasphere DAQ to the CAEN V1740D and a pulser signal was injected into both systems at the start of each run to measure the overall timestamp offset. The \textsc{root} trees were correspondingly combined to form a tree of fission events containing all of the above information about the fragments and the $\gamma$ rays detected in coincidence with them. About $2.5\times 10^9$ fission events were recorded over approximately $180 \text{ h}$ of experiment live time.

\subsection{TFGIC calibration}
We calibrated the TFGIC following the $2E$ method, described in Ref.~\cite{Budtz1987}. Energy- and angle-dependent corrections for energy straggling in the carbon foil backing, TKE-dependent mass corrections for neutron multiplicity~\cite{Gook2014}, and pulse height defect are accounted for in this iterative procedure and calibrated based on previously measured fragment kinetic energy distributions. More details on the chamber and its operation can be found in Ref.~\cite{Marin2023}. The resolutions are as follows: TKE$\sim$ 3-4 MeV FWHM, $\cos \theta_{\text{f}} \sim $ 0.11 FWHM, and $A \sim$ 5-6 AMU FWHM.


\subsection{Gammasphere calibration}

\begin{figure}[htb!]
    \centering
    \includegraphics[width = \figSize\columnwidth]{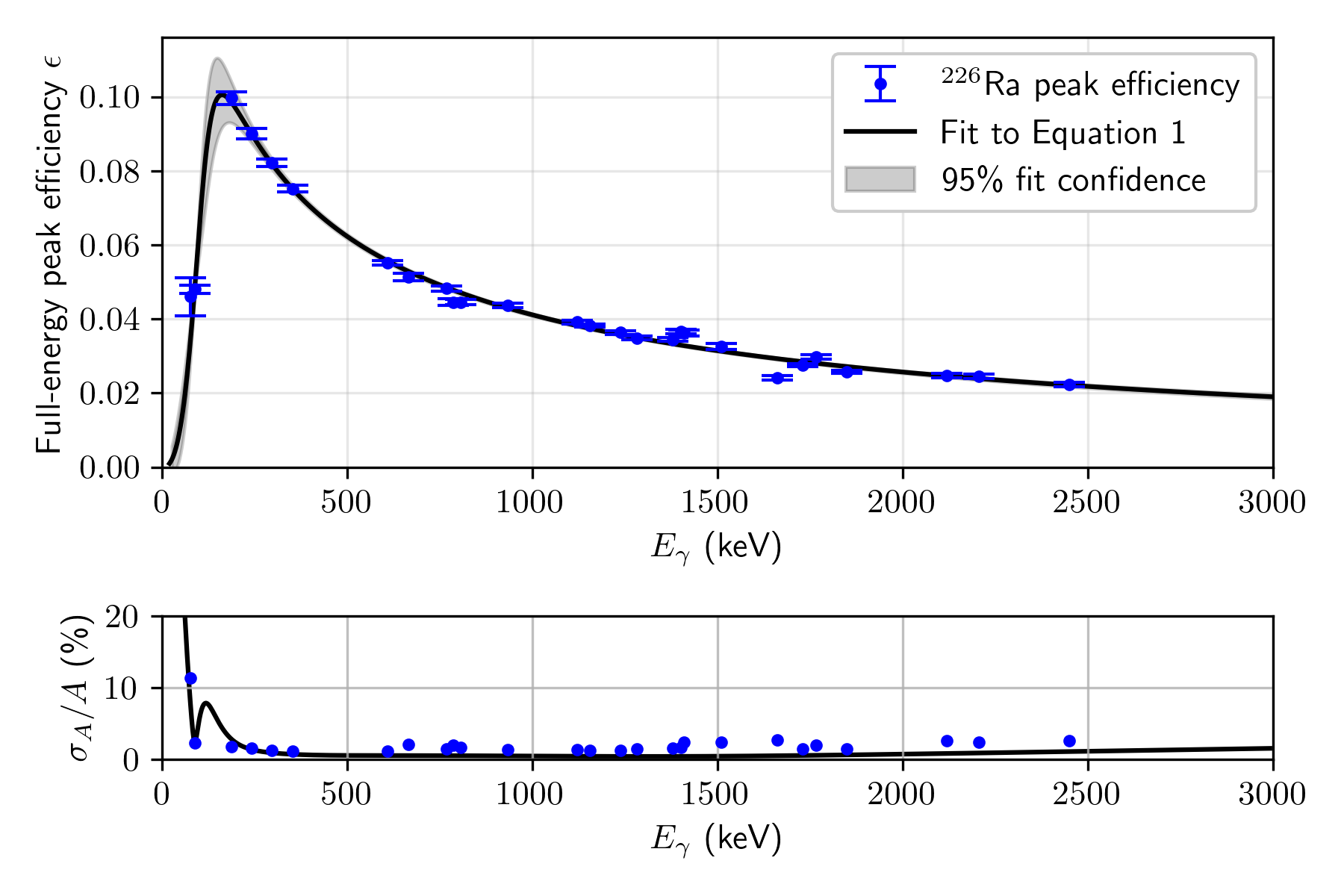}
    \caption{The upper panel shows the measured full-energy peak efficiency of Gammasphere for many $E_\gamma$ and the resulting $\epsilon(E_\gamma)$ fit. The grey band represents the spread of the fit as a function of $E_\gamma$, found by generating many efficiency curves from the fit covariance matrix. The lower panel shows the relative uncertainty (\%) for the measured points and the fit.}
    \label{fig:eff}
\end{figure}

The energy response of each HPGe detector was linearly calibrated with a $^{207}$Bi radioactive source. The full-energy peak efficiency was calibrated by placing a sealed 195.8~kBq $^{226}$Ra source (certified March 1, 2023)\textemdash in equilibrium with its daughters\textemdash inside the TFGIC prior to loading the $^{252}$Cf source. The assembly was placed inside Gammasphere for 30 minutes. The total full-energy peak efficiency for the entire array was determined for energies ranging from the 74.8~keV $K\alpha$ X-ray to the 2447.9-keV $^{214}$Bi $\gamma$ ray, and comparing the measured counts to the expected total emission. The points were fit to Equation~\ref{eq:eff}

\begin{equation}
    \log \epsilon = [(A+Bx)^{-G} + (D+Ey+Fy^2)^{-G}]^{-1/G},
    \label{eq:eff}
\end{equation}
where $x = \log (E_\gamma / \text{100 keV})$ describes the behavior at low $E_\gamma$ and $y = \log (E_\gamma / \text{1000 keV})$ the behavior at high $E_\gamma$, and $G$ the sharpness of the interaction between the regions, from RadWare's EFFIT routine~\cite{radware}. While Gammasphere holds up to 110 HPGe detectors, 63 were placed in the array for this experiment and 54 were used in the final analysis. The others were discarded due to poor resolution or peak-to-total ratios. The total full-energy peak efficiency, $\epsilon (E_\gamma)$, of the array is shown in Fig.~\ref{fig:eff}.

To appropriately correct for the non-uniform angular efficiency, we binned the Gammasphere detectors by $\cos \theta_{C7}$ (the opening angle between each detector and the C7 port\textemdash the location of the HPGe detector aligned with the chamber axis on the ``source" side of the TFGIC). We then repeated the efficiency curve fitting procedure for each angle bin.

\section{Analysis}
\label{sec:Analysis}
The following method was employed to reconstruct the post-statistical spin distribution of $^{144}$Ba, a high-yield fragment (3.37\%~\cite{England1993}) resulting from $^{252}$Cf spontaneous fission, as a function of TKE. This approach is applicable to other fragments for which there are sufficient statistics.

\subsection{Selecting events with $^{144}$Ba}
\label{subsec:Selecting events with $^{144}$Ba}
Several cuts are applied to isolate fission events that contain $^{144}$Ba in the final state. The first cut is a fragment mass cut: fission events for which the reconstructed pre-neutron-emission mass of the heavy fragment, $A_H$, is within 3 u of the post-neutron-emission mass ($A'_H = 144$~u) plus the average neutron multiplicity, $\bar\nu (A, \text{TKE})$ from G\"o\"ok~\textit{et~al.}~\cite{Gook2014}, are accepted~\cite{Marin2023}. The second is a fission axis angle cut, where only fission events that are aligned with the cylindrical axis of TFGIC, $\lvert \cos \theta_\text{f}\rvert > 0.9 $ are accepted. This step is essential to fully constrain the fission axis and perform Doppler correction of the prompt $\gamma$ rays, as the TFGIC does not provide a measurement of the azimuthal angle of the fission axis. All $\gamma$ rays in coincidence with a fission event that passes the mass and angle cuts are Doppler-corrected as if they originated from the heavy fragment (because $^{144}$Ba is heavy). The overall effect is that peaks from heavy fragment transitions are sharpened, and those from light fragments are broadened further. Fission events are tagged for further analysis if they contain a $\gamma$ ray with corrected $E^H_\gamma$ in the heavy fragment frame that overlaps with the first-excited-state transition of $^{144}$Ba at 199~keV. Binning the remaining $\gamma$ rays by their $E_\gamma$, the opening angle they make with the light fragment's direction of travel, $\cos \theta_L$, and the total kinetic energy of the fission event, TKE, we obtain the 3-D histogram that contains all of the necessary information to reconstruct the post-statistical-emission spin distribution of $^{144}$Ba as a function of TKE. Since the fragments are emitted back-to-back in spontaneous fission, the choice of $\cos \theta_L$ instead of $\cos \theta_H$ is arbitrary and does not affect the analysis. Each bin entry is adjusted using $\cos$-dependent efficiency curves to enable the reconstruction of angular distributions of discrete transitions relative to the fission axis.



\subsection{Total kinetic energy binning}
\label{subsec:Total kinetic energy binning}
The 3-D spectrum is sliced by total kinetic energy (TKE) to examine correlations between energy and spin. Selected events in the median 98\% of the TKE distribution are binned such that each TKE bin contains about the same amount of fissions. This step minimizes systematic uncertainties that could arise from bins of uniform width, but non-uniform statistics. There are about $10^5$ fissions in each of the seven TKE bins.

\subsection{Spectrum fitting}
\label{subsec:Spectrum fitting}
Each TKE slice is a 2-D spectrum, with axes $E_\gamma$ and $\cos \theta_L$. Transitions with fixed $E_\gamma$ from $^{144}$Ba, as well as some from its common partner fragments, $^{102,104,106}$Mo, appear as ``tilted'' lines with respect to $\cos \theta_L$. Since the transitions from the light and heavy fragments are Doppler-shifted in opposite directions with respect to $\cos \theta_L$, the two can be separated: the light fragment lines are tilted forward such that a higher lab-frame $E_\gamma$ is observed at large $\cos \theta_L$, while the heavy fragment lines are tilted backward. 

\begin{figure}[htb!]
    \centering
    \includegraphics[width = \figSize\columnwidth]{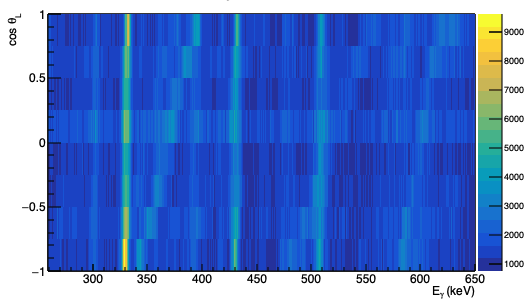}
    \caption{2-D $\gamma$-ray spectrum of $^{144}$Ba, Doppler-corrected to the heavy fragment frame. Vertical lines from $^{144}$Ba are visible at $E_\gamma = 331,431, 509\text{ keV}$. The 369-keV line from $^{104}$Mo is also visible, ranging from $E_\gamma(\cos \theta_L = -1) \approx 346$ keV to $E_\gamma(\cos \theta_L = 1) \approx 395$ keV.}
    \label{fig:2dspecCorr}
\end{figure}

To achieve the best possible $E_\gamma$ resolution for $^{144}$Ba transitions, all measured $\gamma$ rays were Doppler-corrected to the heavy fragment center-of-mass frame. The velocities of the fragments were determined from the mass and TKE, and $\cos \theta_L$ was determined from the opening angle between the TFGIC cylindrical axis and the position of the HPGe detector that registered the count. The result, shown in Fig.~\ref{fig:2dspecCorr}, is vertical $^{144}$Ba lines and further tilted Mo lines.

The $\gamma$-ray background is estimated using a statistics-sensitive non-linear iterative peak-clipping (SNIP) algorithm~\cite{Morhac1997}, implemented in \textsc{root}'s TSpectrum class, and subtracted. The net 2-D spectrum is globally fit with a model containing lines that are gaussian broadened for the discrete $^{144}$Ba and $^{104,106}$Mo transitions. The peak energies in the center-of-mass frame are fixed based on the transition energies in ENSDF~\cite{ENSDF_144Ba, ENSDF_104Mo, ENSDF_106Mo}. Since all $\gamma$ rays have been Doppler-corrected to the heavy fragment frame, an additional $\cos \theta_L$-dependent term is included for the partner fragment peak locations. The peak areas are left to be fit simultaneously, where the former comprises the first three Legendre polynomials with appropriate coefficients $C_0 P_0 + C_1 P_1(\cos \theta_L) + C_2 P_2(\cos \theta_L)$. $C_1 \neq 0$ results from residual Doppler asymmetries, while $C_2 \neq 0$ is necessary in cases where the fragment spin is somehow aligned with respect to the fission axis and $\gamma$ rays are emitted anisotropically (such as stretched $E_2$ $\gamma$ rays). Higher-order $P_n$ are not included to avoid overfitting to few noisy angular bins. Interestingly, the $330.8$-keV $4^+ \rightarrow 2^+$ yrast transition exhibits a $C_2/C_0 \sim 1/3$, indicating a high degree of alignment perpendicular to the fission axis in the substates of the $4^+$ state. The integrals of these peaks, and the uncertainties thereof, are calculated from the fit parameters and used to determine the relative intensities of the transitions.

Because we gate on the $2^+ \rightarrow 0^+$ 199-keV transition, its intensity cannot be measured from this spectrum. We instead perform a similar gating procedure on the most common partner fragment, $^{104}$Mo, so that both the $^{144}$Ba $2^+ \rightarrow 0^+$ transition and the $4^+ \rightarrow 2^+$, $330.8$-keV transition feeding that state can be fit. Their ratio is used to reconstruct the 199-keV transition intensity. The overlapping $4^+ \rightarrow 2^+$ $332.4$-keV transition from $^{146}$Ba was accounted for by measuring the $^{146}$Ba $2^+ \rightarrow 0^+$ transition as well.

\subsection{Intensity balance}
\label{subsec:Intensity balance}
\begin{figure}[htb!]
    \centering
    \includegraphics[width = \figSize\columnwidth]{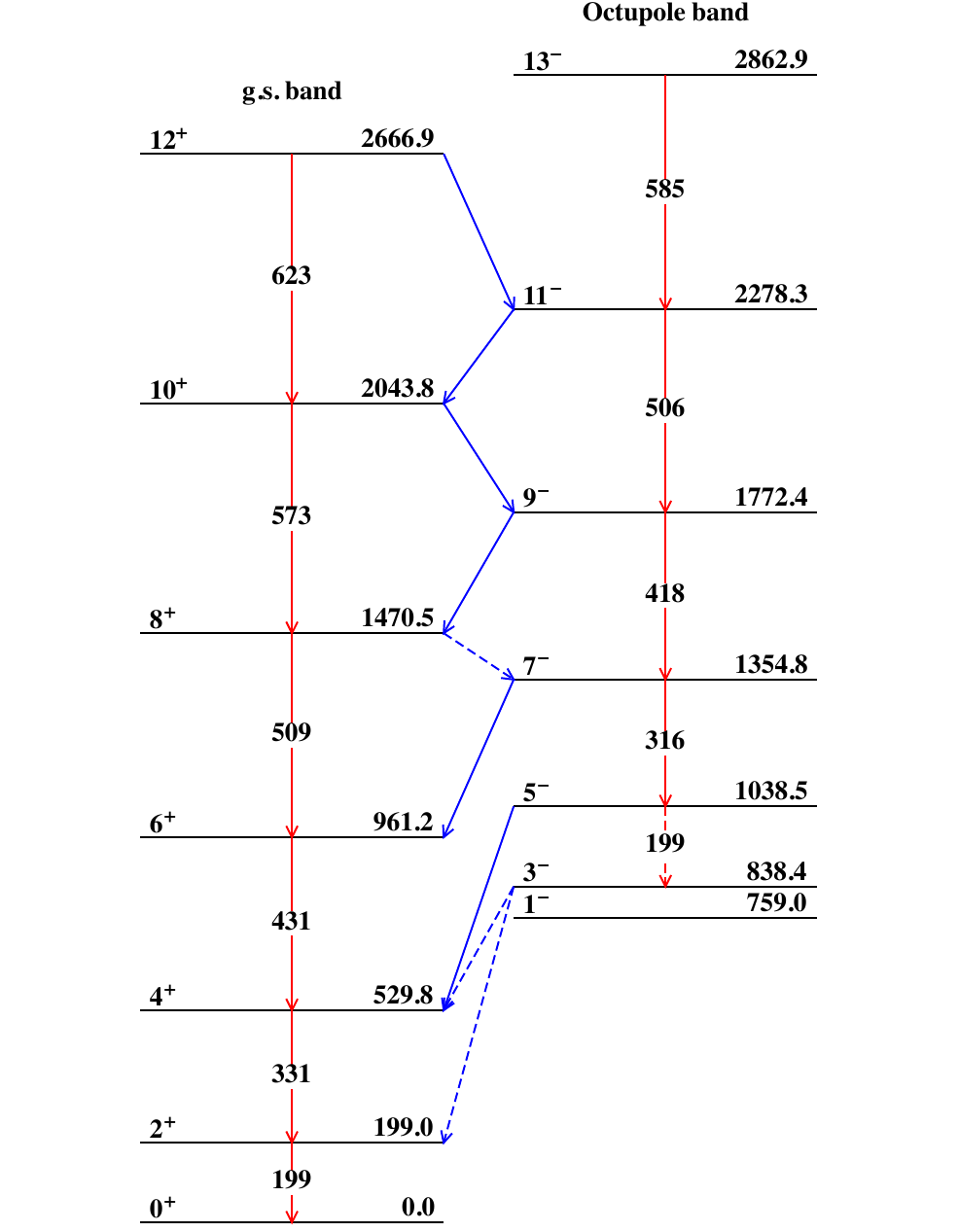}
    \caption{Level scheme of $^{144}$Ba transitions used in the spectrum fit. Level spins and energies, as well as transition energies, were taken from the ENSDF database~\cite{___ENSDF_144Ba}.}
    \label{fig:levelScheme}
\end{figure}

The measured $\gamma$-ray intensities are combined with the well-studied structure of low-lying levels in $^{144}$Ba from the ENSDF database~\cite{ENSDF_144Ba}. Determining the amount of side-feeding $S_i$ from the continuum to a low-lying level $i$, through transitions of known energy, gives direct access to the post-statistical-emission spin distribution of the fragment. The $S_i$ to a level $i$ is simply the sum of all incoming intensities from higher-energy discrete levels subtracted from the sum of all outgoing intensities~\cite{Wilson2021}. The measured $\gamma$-ray intensities are appropriately adjusted for electron-conversion coefficients taken from BrIcc tables with the Frozen Orbital approximation~\cite{Kibedi2008}. Statistical uncertainties from the fitted spectra are propagated through the fit parameters, to the calculated peak intensities. The transitions included in the analysis are shown in Fig.~\ref{fig:levelScheme}. Transitions with solid arrows were observed. Those with dashed arrows, we attempted to fit but they were not observed with any statistical significance. Only ground-state- and octupole-band transitions were observed. While the ENSDF evaluation for $^{252}$Cf(sf) extends to the $16^+$ and $19^-$ states in the ground-state and octupole bands, respectively, those transitions were not observed in this work. The uncertainty for these transitions was assigned based on the intensities of the weakest transitions to which this setup was sensitive.

\subsection{Reconstructing average spin}
\label{subsec:Reconstructing average spin}
\begin{figure}[htb!]
    \centering
    \includegraphics[width = \figSize\columnwidth, trim={0 0 0 0.76cm},clip]{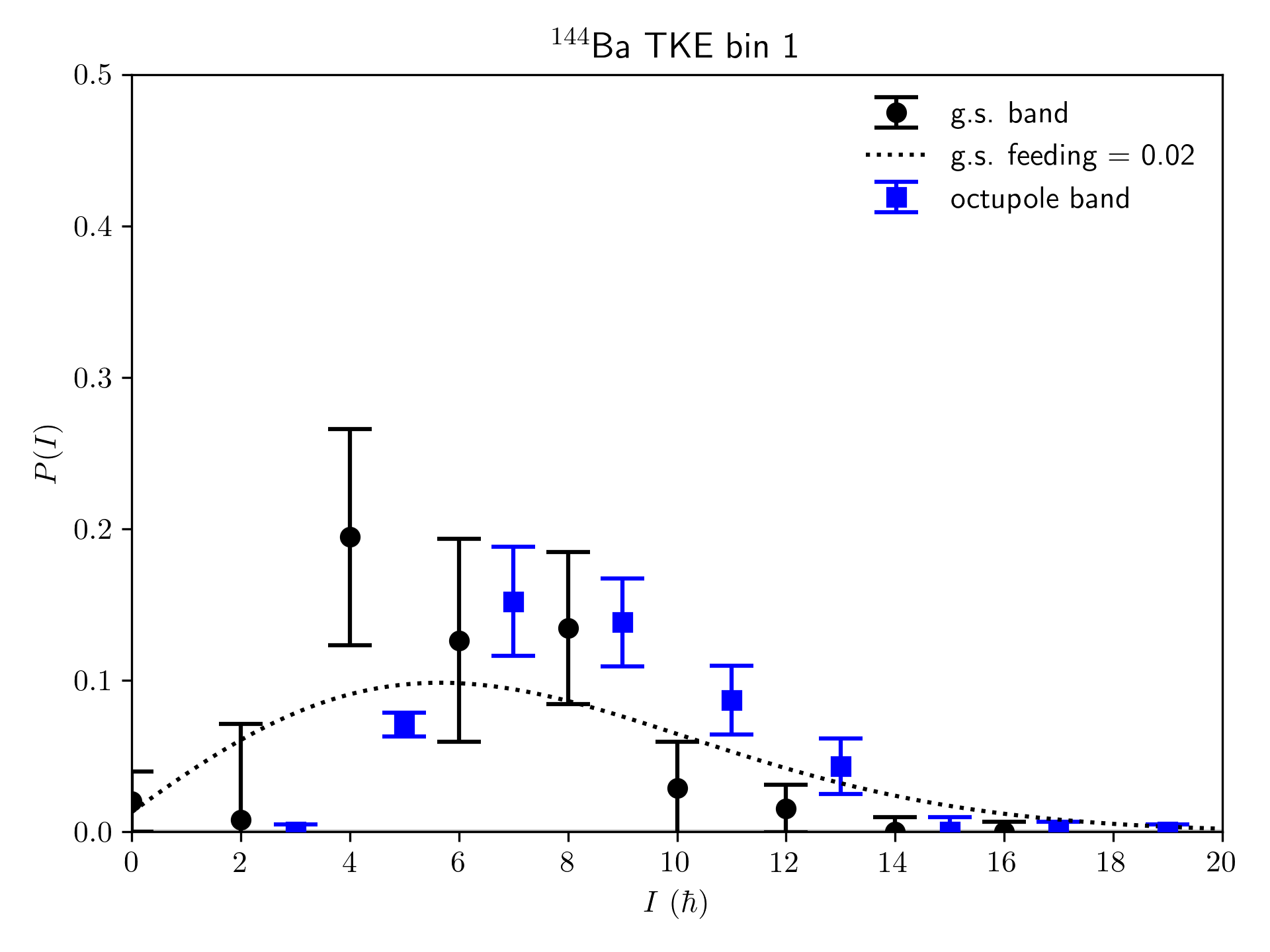}
    \caption{Reconstructed spin feeding distribution for a TKE bin, $173-178\text{ MeV}$. The spin feeding to the ground state band and the octupole vibrational band ($K^\pi = 1^{-1}$) are shown separately. }

    \label{fig:spindist}
\end{figure}

After calculating the side-feeding to each low-lying level, the feedings were correlated to the spin of that level, $I$, and normalized to form a spin probability distribution $P(I)$ like in Fig.~\ref{fig:spindist}. The average of the distribution $\langle I \rangle $ is simply $\sum_i I_i S_i$, where $i$ runs across the low-lying levels.

The statistical uncertainty in the intensities and the uncertainty in the efficiency, which is between 1-5\% for most yrast $\gamma$ rays, were propagated in the following manner to preserve the effect of potentially large covariances in the side-feedings $S_i$. An ensemble of possible intensity ``measurements'' were randomly generated based on (i) their statistical uncertainties and (ii) the covariance matrix of the efficiency curve fit parameters. In other words, each intensity was sampled from a normal distribution, then the efficiency curve was perturbed and the intensities adjusted accordingly. The $S_i$ for all levels were calculated for each generation, such that a covariance matrix $\text{cov} (S_i, S_j)$ could be constructed and used in the calculation of $\langle I \rangle$.

The direct feeding from the quasi-continuum to the ground state is not directly measurable through classical $\gamma$-ray spectroscopy. We estimate the ground-state feeding by fitting the ground-state-band feeding with a statistical spin distribution~\cite{Bethe1936} and extrapolating to $I=0$. The distribution is shown for one TKE bin in Fig.~\ref{fig:spindist}.

\subsection{Potential biases}
\label{subsec:Potential biases}

This analysis introduces some potential sources of bias. Those sources and their effects on the results are discussed here.\\

\textit{Mass and angle cuts}\textemdash
A subset of all measured fission events was chosen based on the fragment mass and angle measurements from the TFGIC. The heavy fragment pre-neutron mass for events with post-neutron $^{144}$Ba is negatively correlated with TKE. Thus a potential acceptance bias toward higher pre-neutron masses could lead to a bias toward lower TKE, or vice versa. This effect is accounted for when setting the mass gate, using experimental average neutron multiplicity data as a function of mass and TKE, or $\bar\nu (A, \text{TKE})$, from Ref.~\cite{Gook2014}. The mass-gate width of 6 u is much larger than the reported uncertainty on $\bar\nu (A, \text{TKE})$, so no bias is expected.

The fragment kinetic energies are corrected for angle-dependent effects, particularly higher energy loss. Since the fission axis has no preferred direction with respect to any quantities measured in the experiment, no bias is expected to arise from the angle cut.

\textit{Background subtraction and fitting}\textemdash
Transitions from $^{144}$Ba, as well as strong transitions from partner fragments and other sources, produce features in the $\gamma$-ray spectra that can be fit simultaneously. However, the background of statistical and weak, unrelated $\gamma$-ray transitions cannot be easily separated and subtracted. This background was estimated with the Background method of \textsc{root}'s TSpectrum class, and parameters were chosen to prevent overfitting. The most important parameter, the clipping window width, was perturbed to assess the sensitivity of the results to the choice of window width. It was found to be relatively insensitive.\\

\textit{Incomplete level scheme knowledge}\textemdash
One potential source is incomplete level-scheme knowledge. Only known $^{144}$Ba transitions in the level scheme were fit, and thus this analysis is insensitive to side-feeding of levels that are absent in the evaluation. This source was considered negligible since all peaks with substantial intensity were identified, and thus the feeding to some unknown state\textemdash and its impact on the reconstructed average spin\textemdash would be insignificant for any reasonable level spin.\\

\textit{Intensity bypassing the first excited state}\textemdash
This method is insensitive to any flow of intensity that bypasses the first-excited state. For $^{144}$Ba, two such paths exist to bypass the 199-keV first excited state: the 758.9~keV ($1^-$) and 1864.2~keV ($2^+$) states have been observed to decay directly to the ground state. Neither of these direct-to-ground-state transitions, however, were observed alongside the 199~keV transition when gating on common partner fragment $^{104}$Mo. We concluded that for $^{144}$Ba, the impact of any missed intensity due to gating on the first-excited state was negligible.\\

\textit{Ground-state feeding estimation}\textemdash
The method for estimating feeding from the statistical regime directly to the ground state described in Section~\ref{subsec:Reconstructing average spin} relies on the assumption that the shape of the real spin probability distribution resembles that of Bethe's work from statistical considerations~\cite{Bethe1936}. This assumption results in an estimation of small ground-state feeding ($\sim$ few \%) in $^{144}$Ba, which is reasonable since $^{144}$Ba is easily deformed. We abscribe a large relative uncertainty to this point due to the model dependence.\\

\textit{Overlapping $\gamma$ rays}\textemdash
In some cases, transitions from the same nucleus are too close in energy to resolve. The total intensity of these $E_\gamma$ multiplets is divided among the transitions in a way that avoids reconstructing unphysical negative side-feedings. The side-feeding distribution is calculated for all possible partitions, and among the allowed partitions, and best estimate is taken as the mean allowed strength. Transitions that are close in energy but originate from a partner fragment or another stationary background source can be separated based on Doppler considerations, so this estimation only needs to be done for $^{144}$Ba lines. \\

\textit{Contamination from other fragments}\textemdash
Potential contamination from heavy fragments other than $^{144}$Ba could arise if that contaminant is i) similar in mass and ii) shares two transition energies with $^{144}$Ba, one being the 199-keV first-excited-state transition. The potential for contamination in the case of $^{144}$Ba was evaluated with  \textsc{fifrelin}~\cite{Litaize2015}, a Monte Carlo code which models the de-excitation of fission fragments. It was determined that no heavy fragments would exhibit appreciable contamination: even the most likely candidates of $^{145}$Ba and $^{147}$La would produce doublets of two to three orders of magnitude weaker intensity, well within statistical uncertainty.

\section{Results}
\label{sec:Results}

\begin{figure}[htb!]
    \centering
    \includegraphics[width = \figSize\columnwidth]{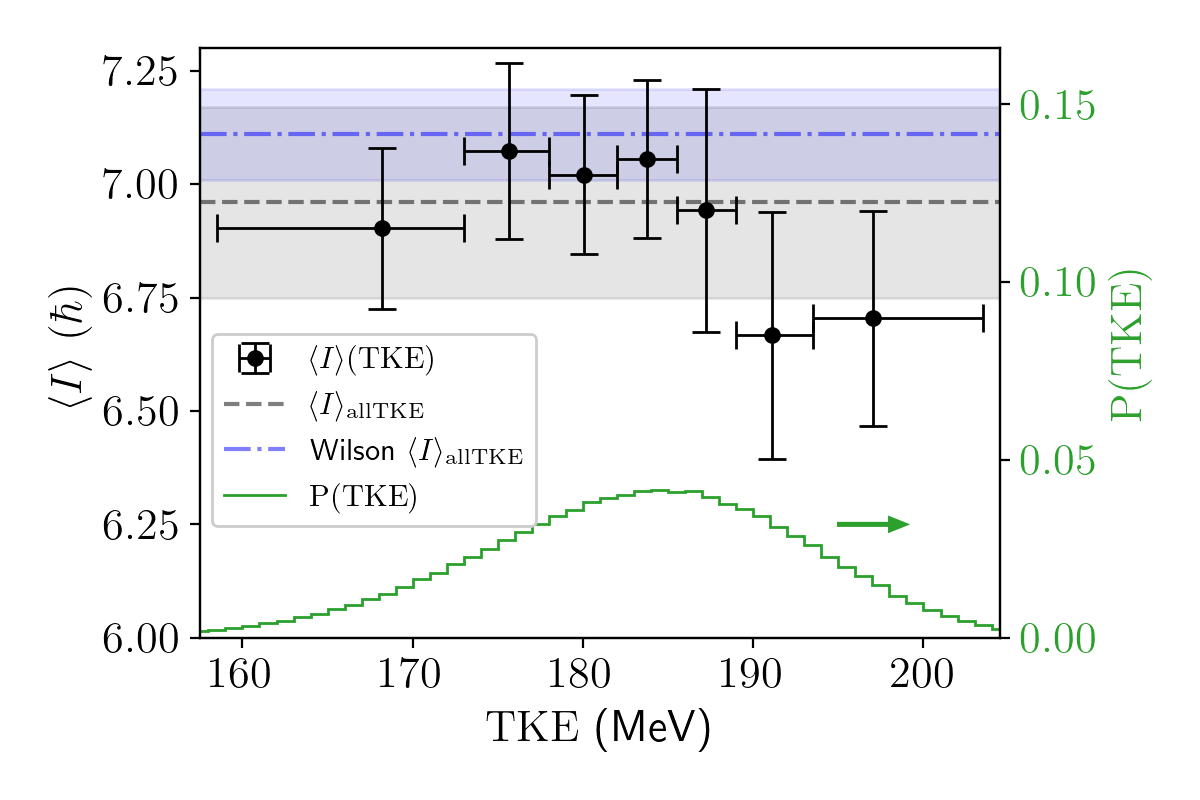}
    \caption{Measured $\langle I \rangle$ as a function of TKE for $^{144}$Ba. The TKE-integrated value of $\langle I \rangle_\text{all} = 6.96\pm 0.21$ (black dashed) agrees well with that of Wilson~\textit{et~al.}~\cite{Wilson2021}, $7.11\pm 0.09$ (post-statistical emission, blue dash-dotted). The horizontal bars signify the widths of the TKE bins. The underlying TKE distribution is shown in green on the right axis.
    }
    \label{fig:tke}
\end{figure}

The average post-statistical spin, $\langle I \rangle$ of $^{144}$Ba was reconstructed for TKE bins with a total range of 158 - 203 MeV (roughly equivalent to a total excitation energy (TXE) range of 12 - 58 MeV). The results are shown in Fig.~\ref{fig:tke}, where the TKE probability distribution (green histogram) is included in the background. $\langle I \rangle$ changes only by about $0.5$ $\hbar$ across the entire TKE range. The uncertainties in $\langle I \rangle$ include the statistical and fit uncertainties of the measured intensities, as well as the efficiency uncertainty, as described in Section~\ref{subsec:Reconstructing average spin}. The horizontal bars denote the widths of the TKE bins, while the horizontal location of the points indicates the average TKE in that bin.

\section{Discussion}
\label{sec:Discussion}
We interpret the minimal dependence of $^{144}$Ba post-statistical $\langle I \rangle$ on TKE as indicative of insensitivity of the spin to this fragment's initial excitation energy.

This interpretation relies on the assumption that the neutron and statistical $\gamma$-ray emission, which precede the measured discrete $\gamma$-ray transitions, do not significantly alter the spin distribution. G\"o\"ok~\textit{et~al.}~\cite{Gook2014} demonstrated experimentally that neutrons are emitted isotropically in the fragment frame in $^{252}$Cf(sf). Hoffman~\cite{Hoffman1964}, Marin~\textit{et~al.}~\cite{Marin2022}, and Val`ski\textit{ et al.}~\cite{Valski1969} each observed that statistical $\gamma$-ray transitions are mostly isotropic. Since fragment spins are generally aligned perpendicular to the fission axis~\cite{Wilhelmy1972}, these results show that neutrons and statistical $\gamma$ rays cannot significantly lower the anisotropy\textemdash and therefore the magnitude\textemdash of the fragment spin without being anisotropic themselves.

For a practical validation of this assumption, Wilson~\textit{et al.}~\cite{Wilson2021} used a similar technique to the one presented in this work to reconstruct the post-statistical $\langle I \rangle$ of fragments as a function of fragment mass, or $\langle I \rangle (A)$, and found variations of several $\hbar$. Any effect of the statistical neutron and $\gamma$-ray emission, if present, on the post-statistical spin measurement is insufficient to destroy large correlations. 

The TKE is complementary to the TXE of the fragments for a given mass split, and since we know from neutron multiplicity measurements~\cite{Gook2014} that $^{144}$Ba receives about half of the total TXE, this behavior indicates that the post-statistical spin of $^{144}$Ba does not change significantly with excitation energy. This result is in agreement with the measurement by Wilhelmy \textit{et al.}~\cite{Wilhelmy1972}, who have also observed a spin-TKE independence, although they were sensitive only to the first few ground-state-band transitions and had three TKE bins. The interpretation is validated in view of a previous $\gamma$-ray multiplicity measurement: the $A=106$ panel in Fig. 6 of Ref.~\cite{Marin2024} shows that the $\gamma$-ray multiplicity of heavy fragments around the mass of $^{144}$Ba is nearly flat as a function of excitation energy, especially in the TXE = 12 - 58 MeV range.

The insensitivity of $\langle I \rangle$ in $^{144}$Ba to energy indicates that the spin generated in fission is not compatible with that is expected from statistical theory alone. In such statistical spin generation models the spin of a fragment is sampled based on the fragment temperature, either after calculating the density of states~\cite{Bethe1936, Talou2021, Litaize2015, Schmidt2016} or by exciting normal modes of rotation of the di-nuclear system~\cite{Randrup2014}. In either case, a higher nuclear temperature corresponds to more energy available for population of higher-spin states, raising the average spin. The resulting average fragment spin $\langle I \rangle $ changes as $ \left( E^* \right)^{1/4}$. This relationship arises from only basic assumptions that (i) rotational energy is approximately proportional to $I^2$, and that (ii) nuclear temperature goes as $T\propto \sqrt{E^*}$\textemdash which are present in fission fragment de-excitation models like \textsc{cgmf}~\cite{Talou2021}, \textsc{fifrelin}~\cite{Litaize2015}, \textsc{freya}~\cite{Vogt2009}, and \textsc{gef}~\cite{Schmidt2016}. Such a dependence should cause an observed change in $\langle I \rangle$ of $\approx 50$\% as the $E^*$ is increased by a factor of about $5$, the range we have examined in this work. The relative change we have observed, approximately $5-10$~\% at most, is not compatible with this rudimentary prediction.

Thus, the fission fragment spin does not arise solely from statistical excitation, but it could still play a role alongside other spin generation mechanisms like those based on fragment deformation~\cite{Marevic2021, Marin2024}, fragment orientation from microscopic models~\cite{Scamps2023}, and Coulomb torque~\cite{bertsch2019reorientation, Scamps2022, Randrup2023_coulomb}.

\section{Conclusion}
\label{sec:Conclusion}
We combined for the first time a twin Frisch-gridded ionization chamber with a world-class $\gamma$-ray spectrometer to measure the average spin of fission fragments produced in $^{252}$Cf(sf) as a function of TKE, or $\langle I \rangle (\text{TKE})$. In $^{144}$Ba, the average spin increased as TKE decreased, but only by about $0.5$ $\hbar$ over the measured TKE range of 158.5 to 203.5 MeV. This result aligns with that of Marin~\textit{et~al.}~\cite{Marin2024}, which showed that the $\gamma$-ray multiplicity\textemdash often used as an indirect measure of fragment spins\textemdash of the heavy fragment for this mass split quickly plateaued with increasing excitation energy. We suggest that this insensitivity to TKE, and therefore to excitation energy for $^{144}$Ba, points toward a more complex spin generation mechanism than that which is included in some modern fragment de-excitation models, which populate the spins of fragments solely based on thermal excitation.

In the near future we will apply the developed method to other fission fragments in this dataset to build a picture of the ``spin-energy surface,'' or the average spin as a function of both TKE and fragment. Upon including more fragments, we expect to reveal whether the spin-energy correlations are sensitive to fragment mass, ground-state deformation, or other properties.\\

\acknowledgments
N.P.G. thanks the Low Energy Technical Support group at the ATLAS facility for making this experiment possible, the developers of \textsc{fifrelin} for providing $^{252}$Cf(sf) events, and D.H. Potterveld for help with navigating the computing resources at ANL. N.P.G., S.M., S.D.C., and S.A.P. were funded in-part by the Consortium for Monitoring Technology and Verification under Department of Energy National Nuclear Security Administration award number DE-NA0003920. This material is based upon work supported by the U.S. Department of Energy, Office of Science, Office of Nuclear Physics, under Contract Number DE-AC02-06CH11357. This research used resources of Argonne National Laboratory's ATLAS facility, which is a DOE Office of Science User Facility. Authors also gratefully acknowledge the use of the Bebop cluster in the Laboratory Computing Resource Center (LCRC) at Argonne National Laboratory.

\bibliography{references}


\end{document}